\documentclass{optica-article}

\journal{opticajournal}
\articletype{Research Article}
\usepackage{lineno}
\begin{document}

\title{Self-similarity-based super-resolution of photoacoustic angiography from hand-drawn doodles}

\author{Yuanzheng Ma,\authormark{1} Wangting Zhou,\authormark{2} Rui Ma,\authormark{3} Sihua Yang,\authormark{3,4} Yansong Tang,\authormark{1} and Xun Guan\authormark{1,*}}% and Yonghong He\authormark{2,3,*}}

\address{\authormark{1}Tsinghua-Berkeley Shenzhen Institute, Tsinghua Shenzhen International Graduate School, Tsinghua University, Shenzhen, 518055, China\\
%\authormark{2}Shenzhen International Graduate School, Tsinghua University, Shenzhen 518055, China\\
\authormark{2}Engineering Research Center of Molecular \& Neuro Imaging of the Ministry of Education, Xidian University, Xi’an, Shaanxi 710126, China\\
\authormark{3}MOE Key Laboratory of Laser Life Science \& Institute of Laser Life Science, College of Biophotonics, South China Normal University, Guangzhou, 510631, China\\
\authormark{4}\href{mailto:yangsh@scnu.edu.cn}{yangsh@scnu.edu.cn}
}

\email{\authormark{*}xun.guan@sz.tsinghua.edu.cn} %% email address is required; see note below about the corresponding author designation



\begin{abstract}
Deep-learning-based super-resolution photoacoustic angiography (PAA) is a powerful tool that restores blood vessel images from under-sampled images to facilitate disease diagnosis. Nonetheless, due to the scarcity of training samples, PAA super-resolution models often exhibit inadequate generalization capabilities, particularly in the context of continuous monitoring tasks. To address this challenge, we propose a novel approach that employs a super-resolution PAA method trained with forged PAA images. We start by generating realistic PAA images of human lips from hand-drawn curves using a diffusion-based image generation model. Subsequently, we train a self-similarity-based super-resolution model with these forged PAA images. Experimental results show that our method outperforms the super-resolution model trained with authentic PAA images in both original-domain and cross-domain tests. Specially, our approach boosts the quality of super-resolution reconstruction using the images forged by the deep learning model, indicating that the collaboration between deep learning models can facilitate generalization, despite limited initial dataset. This approach shows promising potential for exploring zero-shot learning neural networks for vision tasks.
% the self-learning evolving neural networks for vision tasks.
% an infinite number of realistic PAA images could be generated from random hand-drawn doodles to enrich PAA image datasets, pushing the boundary of super-resolution models for restoring high-resolution information from coarse PAA images.
% The average Fr\'echet-inception-distance from authentic to forged photoacoustic images is 96.26, showing a better similarity to authentic human lips images than other photoacoustic datasets. average peak signal-to-noise ratio and structure similarity hit 26.95 and 0.8591 in testing, showing the forged dataset is 
\end{abstract}

%%%%%%%%%%%%%%%%%%%%%%%%%%  body  %%%%%%%%%%%%%%%%%%%%%%%%%%
\section{Introduction}
Photoacoustic angiography (PAA) is a non-labeled vascular imaging method revealing the organ tissue structure with high optical contrast\cite{Hu:10, Errico2015, Reza2018, Nie2014, Lin2022}. For oxygen saturation measurement, blood flow monitoring, and cancerous tumor diagnosis\cite{Wang2004, Gao2017}, high spatial resolution PAA is usually necessary to visualize and measure vascular networks. To achieve high spatial resolution and visualize vascular networks in vivo, a small spatial sampling interval is required. However, high-resolution PAA often requires long imaging time, thus compromising imaging speed. The extremely time-consuming imaging process of PAA seriously impedes its clinical applications, while the real-time monitoring of dynamic events in living organs is even more prohibitively difficult\cite{dispirito2020reconstructing}.

A new strategy speeds up the imaging process by taking low-resolution PAA images and boosting their quality by deep-learning-based super-resolution\cite{Li2022, Yang2021, Davoudi2019}, hence eliminating the speed bottleneck of PAA. Yet, a data-driven super-resolution model usually requires a large training dataset size, which takes prohibitively long time to collect \cite{Godefroy2021}. For example, provided the laser-detector-motor system runs faultlessly at 10 minutes per image\cite{zhou2019optical}, the establishment of a small dataset of human lips involving 10240 PAA images with a resolution of 10$\mu m$ would take 71.1 days, and even the fastest MEMS-based sampling system will still takes 7.11 days\cite{Ma2022}. While a dataset is essential for super-resolution models to learn the relationship between low-resolution and high-resolution PAA images, it also varies over tissues and organs, further increasing the workload of database collection. 

Endeavors are made into artificially enlarging training datasets over real PAA images, so that fewer images are collected in practice. \cite{Godefroy2021} introduced transfer learning to complement training images, and greatly improved image quality without a large size of experimental dataset. With the help of a Bicycle-GAN-based data generator, \cite{Ma2022} realized the PAA dataset enlargement for training the super-resolution PAA model. Some researchers utilized the k-wave-based simulation method to increase the total high-resolution image number for training\cite{Hauptmann2018, Yazdani2021, Tong2020}. While the previous methods tried to reduce the number of images to capture, they broke the consistency between the training and test sets, thus achieving a poor performance of super-resolution models during the inference and low confidence in the reconstructed images\cite{Awasthi2020, Hsu2021}.

This work proposes a method enabling PAA super-resolution with considerably fewer real PAA images. The proposed method comprises two phases, each of which trains one model. In phase 1, a small batch of authentic PAA images is used to train the data generation model, which forges PAA images from hand-drawn doodles with diffusion probabilistic model\cite{Saharia2021}. The diffusion-probabilistic-model-based method is more generalizable: any change in the input doodle or the diffusion process leads to a new image in inference; therefore, this method can produce an infinite number of forged PAA images. This phase differs from general data augmentation techniques such as figure rotation or flipping, as it learns the from limited input patterns and improves the quality and diversity of the training data\cite{kobeissi2022enhancing}. In phase 2, the forged images are employed to train SwinIR for super-resolution PAA, utilizing the self-similarity of the blood vessel structure\cite{Liang2021, Zhang2022SwinFIR}. Our quantitative analysis shows that while testing super-resolution PAA models with these images, forged PAA images also achieve comparable or even better performance compared to authentic PAA images. This, in turn, verifies the verisimilitude of the forged images and the effectiveness of the proposed method. To demonstrate the reconstruction ability of the super-resolution model for the data with the additional time dimension, we verify the dynamic restoration of the model by reconstructing the video-frame sequence of human lips in PAA continuous monitoring. The results reveal a temporally consistent reconstruction from low-resolution video, which paves the way to real-time implementation of PAA applications in the clinic. 

We would refer interested readers to the GUI for generating blood vessel images from doodles in GitHub at https://github.com/yuanzhengthu/handDrawnPAAImages.

\section{Methods}
\subsection{Iterative reinforcement-based PAA image generation}

The image generation model used to transform hand-drawn doodles into corresponding PAA images is based on the super-resolution diffusion probabilistic model. This model iteratively generates a super-resolution blood vessel image from a Gaussian noise image $y_T \sim \mathcal{N}(0, I)$ on the condition of a low-resolution PAA image $x$. In each step of diffusion, the learned conditional transition distributions $p_{\theta}(y_{t-1}|y_{t}, x) = \mathcal{N}_{x, f_{\theta}, t}, t\in[1, T]$, where $y_{t}$ is the noise image at the $t$ step, $f_{\theta}$ represents DeNN\cite{Saharia2021} and $x$ is the low-resolution image. With the $t$ decreased, the generated image is more likely to get close to the super-resolution image\cite{Ho2020,Saharia2021}. With the same imaging system as in \cite{Ma2022}, we collect 321 PAA images of human lips to train the generation model. A dataset of this size may not be sufficient to train a super-resolution model, but it can provide enough data for the image generation model to learn the complex structure and data distribution of PAA images. In this case, a  minimum dataset of 32 images is required for training the generation. Our experiments uses 202 PAA images for training the generation model, while 59 images are reserved for testing and 60 for validation. To link doodles to realistic blood vessel images $\tilde{y_0}$, we first train the image generation model to learn the relationship between low-resolution and $N$-times enlarged high-resolution PAA images ($N$ could be 2, 4, 8, 16...). These low-resolution images are down-sampled from the high-resolution ones in OpenCV\cite{bradski2000opencv}.  After training, the generation model is capable of inferring a super-resolution PAA image from its low-resolution counterpart. However, we are curious about generating more realistic high-resolution PAA images by introducing other types of low-resolution images. To achieve this, we create hand-drawn doodles and add normalized Gaussian noise resembling rain to form the blood vessel's trunk and branch, respectively. These images are then down-sample and up-sample these to generate additional low-resolution images, which are fed to the generation model for image generation.

\begin{figure}[!ht]
\centering\includegraphics[width=13cm]{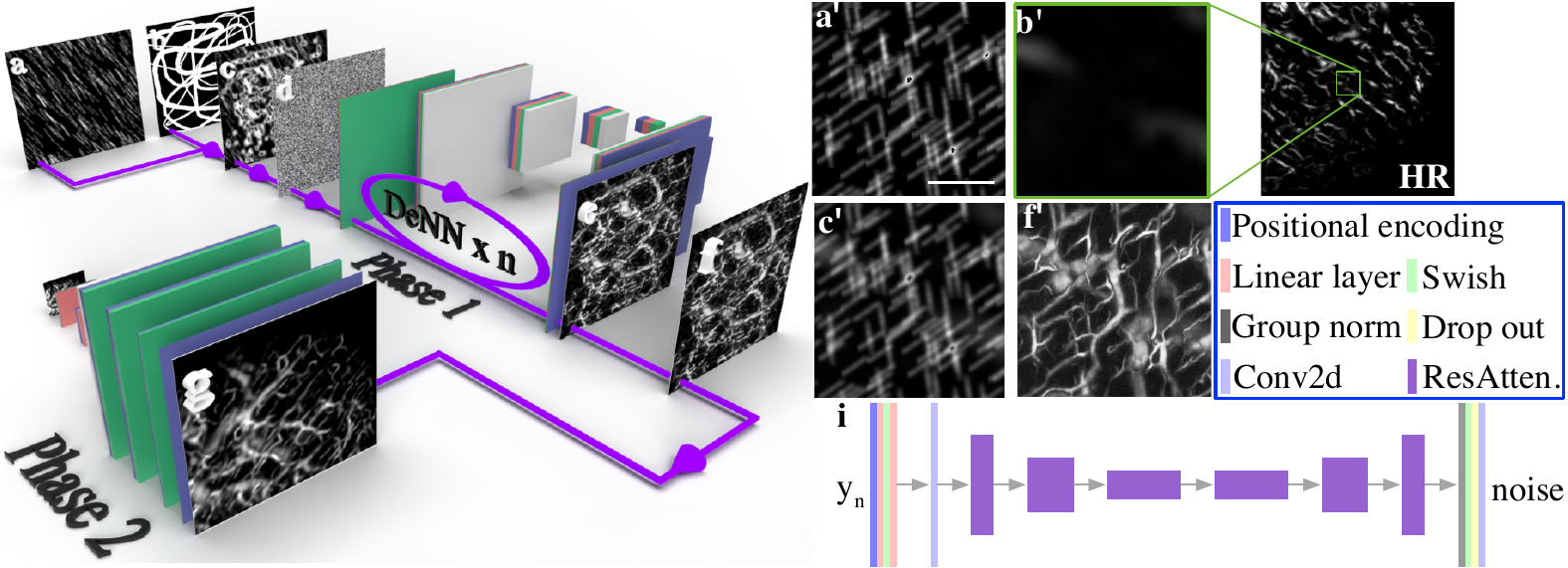}
\caption{(a)-(g) The schematic of training super-resolution blood vessel images for PAA forged from hand-drawn doodles. (a) The rain-like noise image. (b) A hand-drawn doodle. (c) The input of image generation model related to (b) overlapped with (a). (d) The normalized Gaussian noise image. (e) The generated PAA image from (c). (f) The normalized PAA image. (g) The reconstructed super-resolution PAA image. (a')-(f') The schematic of training super-resolution blood vessel images for PAA forged from cropped images, corresponding to the (a)-(f) in the left. (a') The rain-like noise image. (b') A random subset of a high-resolution PAA image \textbf{HR}. (c') The image of (b') overlapped with (a'). (f') The generated PAA image via the generation model from (c'). (i) The structure of UNet-based denoising neural network (DeNN). \textbf{HR}: A high-resolution PAA image. \textbf{ResAtten.}: ResBlock-based attention module. Scale bar: 0.50mm.}
\label{fig1}
\end{figure}

As shown in Fig. \ref{fig1}, in order to deceive the model into identifying the hand-drawn doodle in Fig. \ref{fig1}(b) as a low-resolution photoacoustic image, the doodle is downsampled to its $\frac{1}{N}$ size and upsampled to its original size first. Here $N$ could be 2, 4, 8, 16 or any, but should be consist with the magnification of the super-resolution diffusion probabilistic model. In this experiment, we set $N=8$. Then it is overlaid by the rain-like noise $\epsilon ~ \mathcal{N}(0, I)$, shown in Fig. \ref{fig1}(a). The doodle is blurred with the kernel $[3, 3]$ to finally produce the input of the image generation model $x$, as in Fig. \ref{fig1}(c). To avoid the tremendous work of drawing the doodles piece by piece, we accelerate the PAA image generation process by using upsampled subset of a high-resolution image as the trunk of blood vessel. As illustrated in the Fig. \ref{fig1}, we randomly crop a high-resolution image (marked in Fig. \ref{fig1} as HR) and upsample it to create Fig. \ref{fig1}(b'). We then overlay rain-like noise, as in Fig. \ref{fig1}(a'), on Fig. \ref{fig1}(h'), and blur it with a Gaussian kernel into Fig. \ref{fig1}(c'). Then, Fig. \ref{fig1}(c') is fed into the generation model to produce a realistic high-resolution PAA image, as shown in Fig. \ref{fig1}(f'). 

The diffusion process associated with DeNN $\times$ n is carried out using a UNet-based denoising neural network, as shown in Fig. \ref{fig1}(i). During inference, the trained DeNN model is applied as a series of cascaded chain to gradually reduce the noise of the Gaussian noise image $y_T$. This process is described by the Markov chain, where the input Gaussian noise of the image is gradually decreased on the condition of input $x$ to produce the realistic PAA image $\tilde{y_0}$ \cite{Li2022, Thermodynamics2015, li2022srdiff}. Since the input doodles contain overlapped rain-like noise, a wide range of blood vessel images can be generated by modifying the doodles or the distribution of rain-like noise during the diffusion process. The image reconstruction process can be described as follows:

\begin{equation}
\label{eq1}
y_{t-1} \leftarrow \frac{1}{\sqrt{\alpha_t}}(y_t - \frac{1-\alpha_t}{\sqrt{1-\gamma_t}}f_{\theta}(x,y_{t},\gamma_t)) + \sqrt{1-\alpha_t}\epsilon_t,
\end{equation}
where $y_t$ represents the noise image at step $t$ and $y_0$ is the corresponding high-resolution PAA image. The hyper-parameter $\alpha_t$ is described by an increasing function (see code for details), and $\gamma_t = \prod_{i=1}^t \alpha_i$. The variable $\epsilon_t \sim \mathcal{N}(0,\bf{I})$, and $f_{\theta}(x,y_{t},\gamma_t)$ represents the estimated noise in the image from DeNN. During the process, we treat the noise addition process as adding the output of DeNN to the image, and the denoising process as subtracting the result of DeNN from the image. DeNN can be trained during the noise addition process by gradually degrading the image to Gaussian noise. In this way, the noise addition process is reversed into a denoising process to generate realistic PAA images from the doodles in inference.

\subsection{Automatically screening of high peak signal-to-noise-ratio (PSNR) photoacoustic image with K-means}
\begin{figure}[!ht]
\centering\includegraphics[width=7cm]{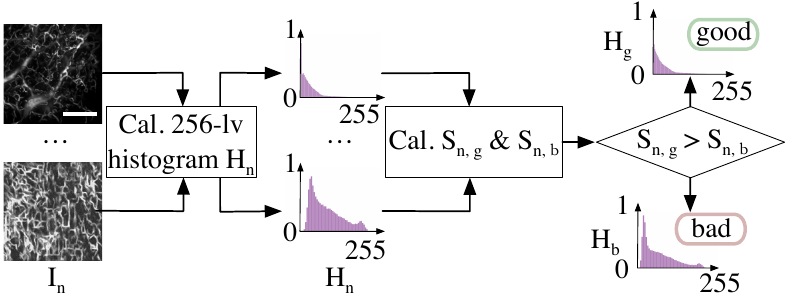}
\caption{The flowchart of K-means clustering algorithm for PAA images screening. $I_n$: $n_{th}$ input PAA image. $H_n$: the histogram of $I_n$. $H_{g} (H_{b})$: the histogram of high-PSNR (low-PSNR) standard, $S_{g} (S_{b})$ correspond the similarities between input $I_n$ and the good standard (the bad standard). Scale bar: 0.50mm.}
\label{fig2}
\end{figure}

Since not all of the generated PAA have good PSNR, a k-means-based screening method is employed to select PAA images with high PSNR\cite{Zhang2019Access}. The images with low PSNR have a noisy background, which means that the number of low-intensity pixels in the low-PSNR images is much larger than that in the high-PSNR images. As shown in the flowchart Fig. \ref{fig2}, 10 of the generated PAA images with high-PSNR and low-PSNR are manually selected as good and bad standards, respectively. Then, the 256-level histogram of them is calculated as $H_{g} = (h_1,h_2,h_3,h_4,...,h_{256})$ for PAA images with high-PSNR and $H_{b} = (l_1,l_2,l_3,l_4,...,l_{256})$ for PAA images with low-PSNR. The 256-levels histogram of an input PAA image $I_n$ is counted as $H_n = (x_{n,1},x_{n,2},x_{n,3},x_{n,4},...,x_{n,256})$. Then, the distances between $H_n$ and $H_{g}$ and between $H_n$ and $H_{b}$ are computed separately. 

As shown in Fig. \ref{fig2}, the input image $I_n$ is classified by the similarity to high-PSNR. The similarity $\mathcal{S}_{n,g}$ between input image and high-PSNR standard has the following format

\begin{equation}
    \label{eq2}
    \mathcal{S}_{n,g} = \frac{1}{\sqrt{\sum\limits_{i=1}^{256}|x_{n,i}-h_i|^2}}.
\end{equation}

Similarly, $\mathcal{S}_{n,b}$ is computed in the same way. As depicted in the flowchart, the input image is classified into $good$ or $bad$, based on its proximity to either set.

\subsection{The self-similarity-based super-resolution model for PAA reconstruction}

Using the high-PSNR PAA images generated, we effectively train the super-resolution model to enhance the quality and resolution of boold vessel images. The fractal structure of blood vessels has been well-documented in previous studies\cite{uahabi2015applications,gabrys2006blood}, and by combining shallow and deep feature extraction with SWinIR, we can exploit this self-similarity to improve the reconstruction quality of blood vessel images\cite{Liang2021,Zhang2022SwinFIR}. By incorporating residual skip connections in the super-resolution model, we can further improve the reconstruction quality\cite{Liang2021}. This approach is particularly significant in PAA since PAA is capable of imaging fine vascular structures of an organism.
%, which is essential in disease diagnosis\cite{cao2019photoacoustic, ahn2021high, yao2011photoacoustic}
We first train the SWinIR with generated PAA images from the previous phase, where we default use data augmentation techniques such as random flipping and rotation. We evaluate the model's perform by using the L1 distance between reconstructed image and the corresponding ground-truth. The loss function specified in the super-resolution PAA reconstruction is experimentally selected as follows
\begin{equation}
    \label{eq3}
    \mathcal{L} = ||I_{SR} - I_{HR} ||_1 + \alpha || w||_1,
\end{equation}
where $I_{SR}$ is the reconstructed super-resolution PAA image, $I_{HR}$ is the ground-truth, $\alpha$ is the hyper-parameter that adjusts the ratio between L1 loss and the regularization term, and $w$ is the weight vector of the super-resolution model. We choose $\alpha=10^{-4}$ in the 4x super-resolution task, and $\alpha=10^{-3}$ in the 8x super-resolution task.

% \subsection{Automatically screening of high-quality photoacoustic image with K-means}
% The forged photoacoustic images are not all good especially when the required number of forged image is big, big enough to be a dataset. As shown in Fig. \ref{fig3}, the histogram of photoacoustic images with high-PSNR are different from the images with low-PSNR. First, some low-PSNR images were manully figured out. We treat the histogram of the images as a $256$ dimensional vector $(c_1,c_2,c_3,c_4,...,c_{256})$, and calculate the average length of every dimensions of the vector, which are regarded as calibrated origin of low-PSNR images in the $R^n$ space. The origin of high-PSNR images was calibrated in the same way. By calculate the distance between the histogram vector of input unknown image ($(x_1,x_2,x_3,x_4,...,x_n)$) and the two calibrated origins, we can figure out the similarity between input image and the low-PSNR and high-PSNR images, respectively. The similarity has the following format:

% \begin{equation}
%     \label{eq6}
%     \mathcal{S} = \frac{\alpha}{\sqrt{\sum\limits_{i=1}^{n}|x_i-c_i|^2}},
% \end{equation}
% where $\alpha$ is constant. 

\section{Results and Discussion}
\subsection{Differentiation of forged and authentic PAA images}
\begin{figure}[!ht]
\centering\includegraphics[width=7cm]{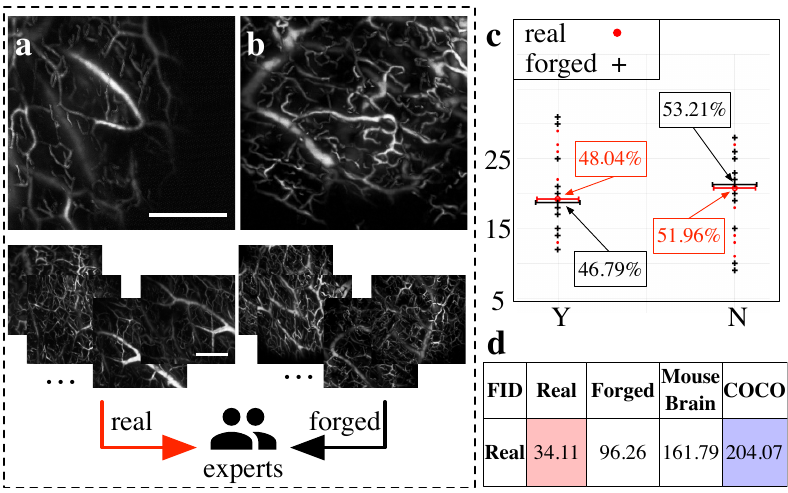}
\caption{The left dashed box encircles the method that 14 experts in photoacoustic imaging for real and forged distinguishing. (a) A real PAA image of human lips. (b) A forged PAA image of human lips. (c) The scatter chart of classifying real and forged PAA images into \textbf{Y} (judged as real PAA images) and \textbf{N} (judged as forged PAA images) by 14 experts in photoacoustic imaging. (d) The \textit{Fr\'echet Inception Distance} scores of real PAA images of human lips (\textbf{Real}), forged PAA images of human lips (\textbf{Forged}), PAA images of mouse brain (\textbf{Mouse Brain}), and COCO dataset (\textbf{COCO}), using real PAA images of human lips as the standard. The longest distance is marked in blue color, while the shortest distance is in red. Scale bar: 0.50mm.}
\label{fig3}
\end{figure}

As shown in Fig. \ref{fig3}, the image of Fig. \ref{fig3}(a) and the images below are the real PAA images of human lips, and the image of Fig. \ref{fig3}(b) and the images below are the forged ones. We invite 14 photoacoustic imaging experts to distinguish between real or forged images using the flowchart presented in the dashed square. We employ a random cropping and shuffling method to create a dataset consisting of 40 PAA images, both authetic and forged. We then sent this dataset to 14 experts and asked them to differentiate between real and forged images. As shown in Fig. \ref{fig3}(c), the labels \textbf{Y} and \textbf{N} indicate whether an image is classified as a real and forged PAA image, respectively. Each red point or black cross represents a result from an expert, and the y-axis refers to the number of samples classified as either \textbf{Y} or \textbf{N}. Our results showed that the probability of authentic PAA images of human lips being classified as real is 48.04\%, while the probability of placing the forged images into the real pool is 46.79\%. The fool rate of the iterative refinement-based PAA image generation is 49.375\%, which is much higher than the fool rate of ESRGAN, which is only 20\%\cite{Hauptmann2018}.

To quantitatively compare the forged PAA images and other images with high-resolution PAA images as the standard, we utilized the \textit{Fr\'echet Inception Distance}\cite{Borji2022}, which measures the distance between two images based on the mean and covariance matrices from the \textit{Inception V3} model. Despite not being trained on authentic PAA images, the \textit{Inception V3} model can still differentiate between blood vessel image sets using the \textit{Fr\'echet Inception Distance}. The \textit{Fr\'echet Inception Distance} between forged and authentic PAA images is

\begin{equation}
    \label{eq4}
    \mathcal{FID}(I,G) = ||\mu_I - \mu_G||_2^2 + Tr(\Sigma_I + \Sigma_G - 2(\Sigma_I\Sigma_G)^{\frac{1}{2}}),
\end{equation}
where $I, G$ stand for the input image and the ground-truth, $\mu_I, \mu_G$ are mean matrices of the input and the ground-truth, $Tr()$ is the trace of a matrix, and $\Sigma_I, \Sigma_G$ are covariance matrices of the input and ground-truth, respectively. The results of \textit{Fr\'echet Inception Distance} between different datasets are listed in Fig. \ref{fig3}(d), showing the distance from authentic to forged PAA images is much shorter than the distance to others. A distance of 34.11 exists between the authentic PAA image set and the 10 manually selected PAA images. This is because 10 manually selected PAA images could not be more significant to represent the whole dataset. Clearly, the forged images are much similar to the real images than mouse brain and COCO, as indicated by smaller FIR values. These forged PAA images are screened by the k-means-based method and collected for super-resolution training. 
%After K-means clustering, the number of forged PAA images is 1281 or 13277, prepared as the training set for super-resolution model in phase 2.

\subsection{Evaluation on super-resolution reconstruction of under-sampled PAA images}
\begin{figure}[!ht]
\centering\includegraphics[width=13cm]{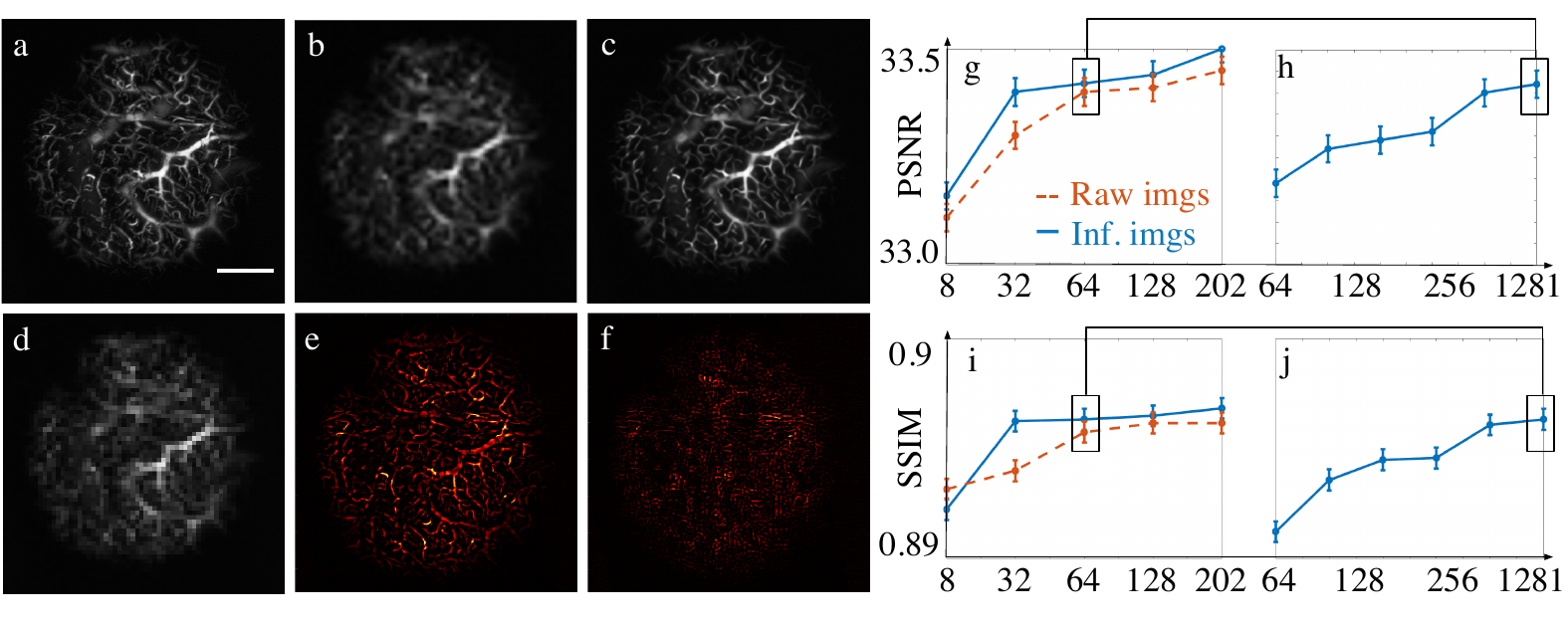}
\caption{The comparison of reconstruction abilities of the super-resolution models trained with different numbers of authentic images. (a) The ground-truth image. (b) The reconstructed high-resolution image using Bicubic algorithm. (c) The reconstructed super-resolution image from our method trained from $202$ authentic images. (d) The low-resolution image. (e) The difference between Bicubic reconstructed image and the ground-truth. (f) The difference between reconstructed image from our method with $202$ authentic images and the ground-truth.  (g)(i) The PSNR and SSIM of the super-resolution model trained with 1281 forged PAA images, whose forgery generation models are trained with a varied number of authentic images, from 8 to 202. (h)(j) The PSNR and SSIM of the super-resolution model trained with a varied number of forged PAA images, whose forgery generation model is trained with 64 authentic images. Connected boxes denote identical data point in (g)(h) for PSNR, and (i)(j) for SSIM. Scale bar: 0.50mm.}
\label{fig4}
\end{figure}

%We use different numbers of PAA images to directly train the super-resolution model as comparison. The results indicate that the forged images could surpass equivalent real images under the structure similarity (SSIM) and peak signal-to-noise ratio (PSNR) metrics.This result suggests that one deep neural network could provide additional information for another. To test the proposed method's  generalization, we use an additional 30 PAA images with distribution different from the training dataset.
We quantify the proposed method's generalization capability by measuring its PSNR and structural similarity index measure (SSIM) with regard to two variables, as shown in Fig. \ref{fig4}. In our experiments, we showcase the efficacy of our super-resolution model in reconstructing high-quality blood vessel images from their 4x down-sampled versions.

We firstly vary the number of authentic images used in the training of the image generation model, in the range of [8, 32, 64, 128, 202]. 1281 images are forged with all these image generation models to train the super-resolution models, while SSIM and PSNR are calculated for the models based on different numbers of authentic images. In comparison, the same number of authentic images are also used to directly train the super-resolution model without image forgery. As shown in Fig. \ref{fig4}(g) and (i), PSNR and SSIM increase with the number of authentic images, with or without image forging model. According to \cite{ajiboye2015evaluating}, a large dataset with good quality can lead to better results when a model has enough parameters for feature extraction. In our experiments, we find that the PSNR of super-resolution from forged images consistently surpass that of authentic images. This improvement in model performance is attributed to the generation model's ability to augment limited input patterns by increasing the number of high-quality PAA images used for training\cite{kobeissi2022enhancing}. However, in the case where only 8 authentic PAA images are available, the SSIM of image forgery is lower. Although a dataset size of 8 is not sufficient to train the generation model, it becomes feasible when the dataset size increases to 32.

Then we fix the authentic image number in image generation model training to 64, and change the number of forged images generated. The results are shown in Fig. \ref{fig4}(h) and (j). With the increase of forged image number from 32 to 1281, the PSNR and SSIM continuously increase. In addition, we screen the generated PAA images before incorporating them into the training dataset to ensure their quality. This approach helps to increase the number of PAA images with good quality for training, which results in improved reconstruction results.

\subsection{Test on super-resolution reconstruction of under-sampled PAA images from other PAA dataset of human lips}

\begin{figure}[!ht]
\centering\includegraphics[width=7cm]{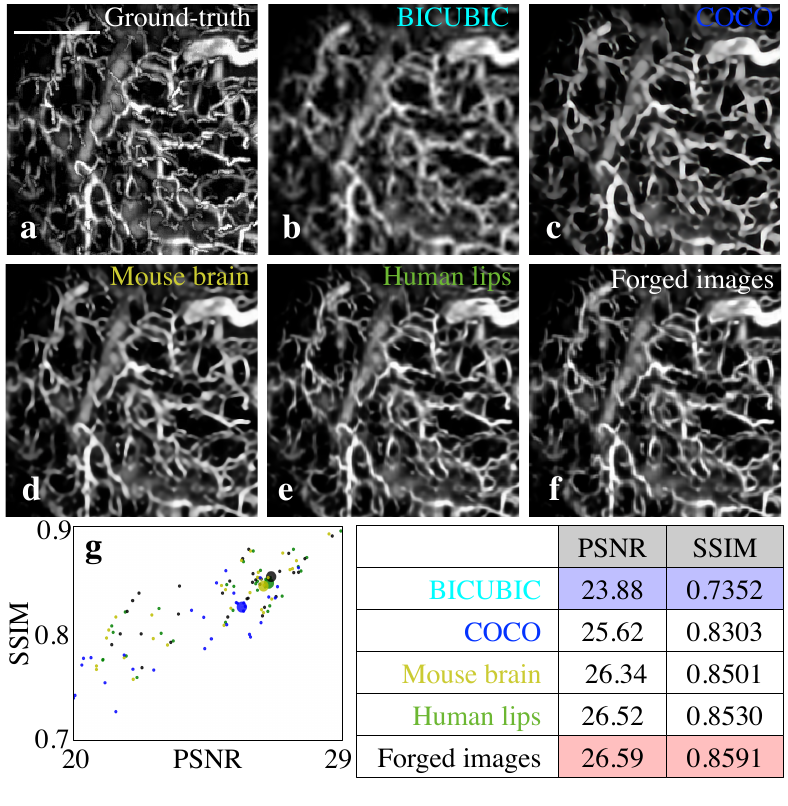}
\caption{Comparison of different reconstructed methods. (a) The high-resolution image of human lips. (b) The image reconstructed by Bicubic algorithm. (c)-(f) The results reconstructed by the super-resolution model trained with (c) COCO dataset; (d) mouse back PAA images; (e) real human lips PAA images; (f) forged human lips PAA images. (g) PSNR and SSIM of different reconstructions, and the mean scores are marked with larger circles; The lower-right chart shows PSNR and SSIM of results from different reconstruction methods, where different datasets are marked with different colors. Scale bar: 0.50mm.}
\label{fig5}
\end{figure}

To further examine the model's generalization ability to reconstruct PAA images of human lips from different dataset and assess its generalization performance, we randomly select 30 PAA images from the dataset in \cite{Ma2022}, and reconstruct their super-resolution from different training datasets, including the COCO dataset\cite{DBLP:journals/corr/LinMBHPRDZ14}, PAA images of mouse brain\cite{dispirito2020reconstructing}, and authentic PAA images of the human lips. Here to train our super-resolution model, we utilize 13277 generated images of human lips, along with 202 authentic images. Apparent differences exist among Fig. \ref{fig5}(a)-(f) even though the super-resolution model and hyper-parameters of training are identical, illuminating the importance of the dataset consistence for super-resolution PAA.  Particularly, the visual reconstruction  of the forged images of human lips is no worse than the authentic PAA images, further validating the generalization capability of super-resolution model trained by the forged images. 

The scatter plot and table in Fig. \ref{fig5}(g) quantifies this generalization capability  in Fig. \ref{fig5}(g). The scatter plot displays the quality indexes of 30 reconstructed PAA images reconstructed from different datasets, labeled with dots in various colors corresponding to the text colors in the table. The mean scores of different datasets are highlighted with dots larger than those of a single image in different colors. As the results obtrained from Bicubic interpolation cannot be compared to those obtrained through other methods, they are not included in this scatter plot. Our results show that the model trained with forged PAA images of the human lips outperforms the model trained with the other datasets in test, even better than the authentic PAA images. This is mainly because the generation model augments the dataset, increasing the variety of blood vessel images. Additionally, the forged PAA dataset, comprising over 10,000 images after screening, is much larger than the authentic PAA dataset with only 202 images.  It is worth noting that the PSNR and SSIM metrics in the table in Fig. \ref{fig5} are lower than those shown in Fig. \ref{fig4}(g)-(h) due to the uncontrollable variations in sampling conditions for the two datasets, such as differences in laser intensity, spot diameter, and sampling interval.

Our experiment confirms the proposed method's capability to reconstruct images of organs, particularly in the senarios where the training dataset is insufficient. In clinical settings, PAA imaging may expose organs to high laser energy, potentially causing damage. However, our proposed method can enhance the resolution of deep organ images yet minimizes the laser exposure.

\subsection{Evaluation on PAA image sequence with video-frame rate}
Dynamic monitoring in vivo needs high-resolution PAA images at a reasonably fast imaging speed, which is infeasible by traditional MEMS-based PAA methods. We future reconstruct PAA images over time to demonstrate the coherence of adjacent frames, which is the key for dynamic monitoring. To accelerate the sampling process, we directly record the maximum projection images of the organ. The high-resolution PAA images is accelerated from $\frac{1}{180}$ frames per second to $\frac{1}{4}$ frames per second. 

\begin{figure}[!ht]
\centering\includegraphics[width=8cm]{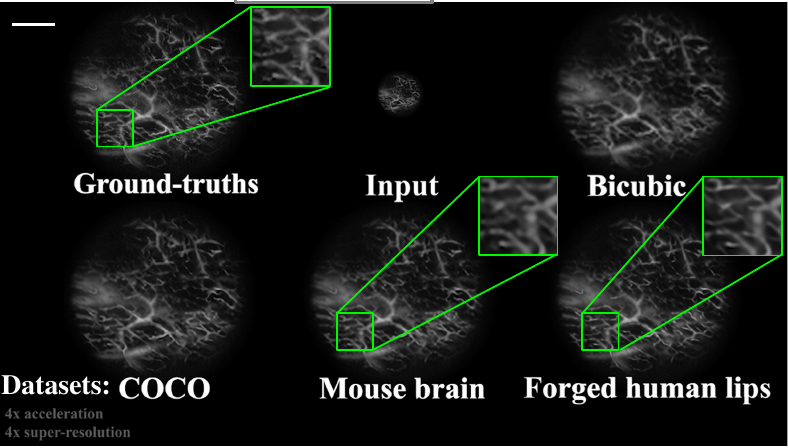}
\caption{The reconstruction frame of under-sampled dynamic PAA imaging of human lips. Scale bar: 0.50mm (see Visualization 1).}
\label{fig6}
\end{figure}

To obtain low-resolution PAA images in a video-frame-rate, we perform a spatial downsample of the original video comprising 33 frames, using a downsample ratio of 1/4, which corresponds to the magnification of our super-resolution model. Subsequently, we utilize the aforementioned methods to reconstruct the dynamic monitoring video. As shown in Fig. \ref{fig6}, the green squares highlight the differences among the ground truths, the results of the super-resolution model trained with mouse brain dataset, and the forged human lips dataset, where the results of the model trained with forged human lips dataset are better than the others. Besides, the results of the super-resolution model trained with forged images of human lips have a cleaner background than the high-resolution sequence, which means the image noise is also removed during the under-sampling reconstruction process. The reconstructed video can theoretically monitor the organ with a 16-factor acceleration in time, as only one forth sampling points are collected in each of the 2 dimensions. This  verifies that super-resolution eliminates the imaging speed bottleneck of PAA.

\section{Conclusion}
%This work proposes a method of deep-learning-based super-resolution PAA imaging that trains the super-resolution model with forged PAA images. 
Deep-learning-based super-resolution methods improve the spatial resolution of PAA while maintaining its temporal resolution. Nevertheless, most existing super-resolution models for PAA are trained with a sizable dataset including at least thousands of authentic blood vessel images, which take an exorbitantly long time to capture. Apart from the burdensome clinical work, the target organ also gets injured upon longstanding laser power exposure. To overcome this limitation, we train a super-resolution model using hand-drawn blood vessel images, and achieve better results than other models, including the one trained on real PAA images. This method alleviates the excessive workload for scanning-based image collection and expands the clinical application of PAA to fast high-resolution PAA imaging or even live video of tissues. In addition, our findings show that the super-resolution model trained on the forged images generated by the generation model surpasses the one trained on real PAA images. This indicates that the iterative training can significantly enhance the model's ability to produce highly realistic images that closely resemble ground-truth images, making it a promising tool for exploring the super-resolution in absence or in short of authentic training data.
%indicating that the iterative training of the generation model with the image generated by itself can enhance its ability to produce highly realistic images, helpful in exploring the super-resolution without any data for training.
%Since high undersampled rates induce uncertainty in reconstructing the PAA image, our following works will focus on reducing the underdetermination during the super-resolution process.

\begin{backmatter}
\bmsection{Funding}
 Shenzhen Science, Technology and Innovation Commission: Fundamental Research Scheme General Program (JCYJ20220530142809022); Stable Support (WDZC20220811170401001). National Key R\&D Program of China (2021YFA1001000).

\bmsection{Disclosures}
The authors declare no conflicts of interest.

\bmsection{Data Availability Statement}
Data underlying the results presented in this paper are not publicly available at this time but may be obtained from the authors upon reasonable request.

\bmsection{Code availability}
The codes, models, and other testing materials related to this paper are available online and can be downloaded from https://github.com/yuanzhengthu/handDrawnPAAImages.

% \bmsection{Supplemental document}
% See Supplement 1 for supporting content. 

\end{backmatter}

%%%%%%%%%%%%%%%%%%%%%%% References %%%%%%%%%%%%%%%%%%%%%%%%%

%%%%%%%%%% If using BibTeX:
\bibliography{optica}

\begin{thebibliography}{10}
\newcommand{\enquote}[1]{``#1''}

\bibitem{Hu:10}
S.~Hu, B.~Rao, K.~Maslov, and L.~V. Wang, \enquote{Label-free photoacoustic
  ophthalmic angiography,} {\protect\JournalTitle{Opt. Lett.}} \textbf{35},
  1--3 (2010).

\bibitem{Errico2015}
C.~Errico, J.~Pierre, S.~Pezet, Y.~Desailly, Z.~Lenkei, O.~Couture, and
  M.~Tanter, \enquote{{Ultrafast ultrasound localization microscopy for deep
  super-resolution vascular imaging},} {\protect\JournalTitle{Nature}}
  \textbf{527}, 499--502 (2015).

\bibitem{Reza2018}
P.~H. Reza, K.~Bell, W.~Shi, J.~Shapiro, and R.~J. Zemp, \enquote{{Deep
  non-contact photoacoustic initial pressure imaging},}
  {\protect\JournalTitle{Optica}} \textbf{5}, 814 (2018).

\bibitem{Nie2014}
L.~Nie, S.~Wang, X.~Wang, P.~Rong, Y.~Ma, G.~Liu, P.~Huang, G.~Lu, and X.~Chen,
  \enquote{{In vivo volumetric photoacoustic molecular angiography and
  therapeutic monitoring with targeted plasmonic nanostars},}
  {\protect\JournalTitle{Small}} \textbf{10}, 1585--1593 (2014).

\bibitem{Lin2022}
L.~Lin and L.~V. Wang, \enquote{{The emerging role of photoacoustic imaging in
  clinical oncology},} {\protect\JournalTitle{Nature Reviews Clinical
  Oncology}} \textbf{19}, 365--384 (2022).

\bibitem{Wang2004}
L.~V. Wang, X.~Wang, G.~Ku, X.~Xie, and G.~Stoica, \enquote{{High-resolution
  photoacoustic tomography},} {\protect\JournalTitle{Conference Proceedings -
  Lasers and Electro-Optics Society Annual Meeting-LEOS}} \textbf{2}, 767--768
  (2004).

\bibitem{Gao2017}
F.~Gao, X.~Feng, R.~Zhang, S.~Liu, R.~Ding, R.~Kishor, and Y.~Zheng,
  \enquote{{Single laser pulse generates dual photoacoustic signals for
  differential contrast photoacoustic imaging},}
  {\protect\JournalTitle{Scientific Reports}} \textbf{7}, 1--12 (2017).

\bibitem{dispirito2020reconstructing}
A.~DiSpirito, D.~Li, T.~Vu, M.~Chen, D.~Zhang, J.~Luo, R.~Horstmeyer, and
  J.~Yao, \enquote{Reconstructing undersampled photoacoustic microscopy images
  using deep learning,} {\protect\JournalTitle{IEEE transactions on medical
  imaging}} \textbf{40}, 562--570 (2020).

\bibitem{Li2022}
J.~Li, C.~Wang, T.~Chen, T.~Lu, S.~Li, B.~Sun, F.~Gao, and V.~Ntziachristos,
  \enquote{{Deep learning-based quantitative optoacoustic tomography of deep
  tissues in the absence of labeled experimental data},}
  {\protect\JournalTitle{Optica}} \textbf{9}, 32 (2022).

\bibitem{Yang2021}
C.~Yang, H.~Lan, F.~Gao, and F.~Gao, \enquote{{Review of deep learning for
  photoacoustic imaging},}  (2021).

\bibitem{Davoudi2019}
N.~Davoudi, X.~L. De{\'{a}}n-Ben, and D.~Razansky, \enquote{{Deep learning
  optoacoustic tomography with sparse data},} {\protect\JournalTitle{Nature
  Machine Intelligence}} \textbf{1}, 453--460 (2019).

\bibitem{Godefroy2021}
G.~Godefroy, B.~Arnal, and E.~Bossy, \enquote{{Compensating for visibility
  artefacts in photoacoustic imaging with a deep learning approach providing
  prediction uncertainties},} {\protect\JournalTitle{Photoacoustics}}
  \textbf{21}, 100218 (2021).

\bibitem{zhou2019optical}
W.~Zhou, Z.~Chen, Q.~Zhou, and D.~Xing, \enquote{Optical biopsy of melanoma and
  basal cell carcinoma progression by noncontact photoacoustic and optical
  coherence tomography: In vivo multi-parametric characterizing tumor
  microenvironment,} {\protect\JournalTitle{IEEE transactions on medical
  imaging}} \textbf{39}, 1967--1974 (2019).

\bibitem{Ma2022}
Y.~Ma, K.~Xiong, X.~Hou, W.~Zhang, X.~Chen, L.~Li, and S.~Yang,
  \enquote{{Cascade neural approximating for few-shot super-resolution
  photoacoustic angiography},} {\protect\JournalTitle{Applied Physics Letters}}
  \textbf{121} (2022).

\bibitem{Hauptmann2018}
A.~Hauptmann, F.~Lucka, M.~Betcke, N.~Huynh, J.~Adler, B.~Cox, P.~Beard,
  S.~Ourselin, and S.~Arridge, \enquote{{Model-Based Learning for Accelerated,
  Limited-View 3-D Photoacoustic Tomography},} {\protect\JournalTitle{IEEE
  Transactions on Medical Imaging}} \textbf{37}, 1382--1393 (2018).

\bibitem{Yazdani2021}
A.~Yazdani, S.~Agrawal, K.~Johnstonbaugh, S.~R. Kothapalli, and V.~Monga,
  \enquote{{Simultaneous Denoising and Localization Network for Photoacoustic
  Target Localization},} {\protect\JournalTitle{IEEE Transactions on Medical
  Imaging}} \textbf{40}, 2367--2379 (2021).

\bibitem{Tong2020}
T.~Tong, W.~Huang, K.~Wang, Z.~He, L.~Yin, X.~Yang, S.~Zhang, and J.~Tian,
  \enquote{{Domain Transform Network for Photoacoustic Tomography from
  Limited-view and Sparsely Sampled Data},}
  {\protect\JournalTitle{Photoacoustics}} \textbf{19}, 100190 (2020).

\bibitem{Awasthi2020}
N.~Awasthi, G.~Jain, S.~K. Kalva, M.~Pramanik, and P.~K. Yalavarthy,
  \enquote{{Deep Neural Network-Based Sinogram Super-Resolution and Bandwidth
  Enhancement for Limited-Data Photoacoustic Tomography},}
  {\protect\JournalTitle{IEEE Transactions on Ultrasonics, Ferroelectrics, and
  Frequency Control}} \textbf{67}, 2660--2673 (2020).

\bibitem{Hsu2021}
K.~T. Hsu, S.~Guan, and P.~V. Chitnis, \enquote{{Comparing Deep Learning
  Frameworks for Photoacoustic Tomography Image Reconstruction},}
  {\protect\JournalTitle{Photoacoustics}} \textbf{23}, 100271 (2021).

\bibitem{Saharia2021}
C.~Saharia, J.~Ho, W.~Chan, T.~Salimans, D.~J. Fleet, and M.~Norouzi,
  \enquote{{Image Super-Resolution via Iterative Refinement},}
  {\protect\JournalTitle{IEEE Transactions on Pattern Analysis and Machine
  Intelligence}} \textbf{PP}, 1--14 (2021).

\bibitem{kobeissi2022enhancing}
H.~Kobeissi, S.~Mohammadzadeh, and E.~Lejeune, \enquote{Enhancing mechanical
  metamodels with a generative model-based augmented training dataset,}
  {\protect\JournalTitle{Journal of Biomechanical Engineering}} \textbf{144},
  121002 (2022).

\bibitem{Liang2021}
J.~Liang, J.~Cao, G.~Sun, K.~Zhang, L.~{Van Gool}, and R.~Timofte,
  \enquote{{SwinIR: Image Restoration Using Swin Transformer},} in
  \emph{Proceedings of the IEEE International Conference on Computer Vision,}
  vol. 2021-Octob (2021), pp. 1833--1844.

\bibitem{Zhang2022SwinFIR}
D.~Zhang, F.~Huang, S.~Liu, X.~Wang, and Z.~Jin, \enquote{Swinfir: Revisiting
  the swinir with fast fourier convolution and improved training for image
  super-resolution,}  (2022).

\bibitem{Ho2020}
J.~Ho, A.~Jain, and P.~Abbeel, \enquote{{Denoising diffusion probabilistic
  models},} {\protect\JournalTitle{Advances in Neural Information Processing
  Systems}} \textbf{2020-December}, 1--25 (2020).

\bibitem{bradski2000opencv}
G.~Bradski, \enquote{The opencv library.} {\protect\JournalTitle{Dr. Dobb's
  Journal: Software Tools for the Professional Programmer}} \textbf{25},
  120--123 (2000).

\bibitem{Thermodynamics2015}
J.~Sohl-Dickstein, E.~A. Weiss, N.~Maheswaranathan, and S.~Ganguli,
  \enquote{{Deep unsupervised learning using nonequilibrium thermodynamics},}
  in \emph{32nd International Conference on Machine Learning, ICML 2015,}
  vol.~3 (2015), pp. 2246--2255.

\bibitem{li2022srdiff}
H.~Li, Y.~Yang, M.~Chang, S.~Chen, H.~Feng, Z.~Xu, Q.~Li, and Y.~Chen,
  \enquote{Srdiff: Single image super-resolution with diffusion probabilistic
  models,} {\protect\JournalTitle{Neurocomputing}} \textbf{479}, 47--59 (2022).

\bibitem{Zhang2019Access}
J.~Zhang, B.~Chen, M.~Zhou, H.~Lan, and F.~Gao, \enquote{{Photoacoustic Image
  Classification and Segmentation of Breast Cancer: A Feasibility Study},}
  {\protect\JournalTitle{IEEE Access}} \textbf{7}, 5457--5466 (2019).

\bibitem{uahabi2015applications}
K.~L. Uahabi and M.~Atounti, \enquote{Applications of fractals in medicine,}
  {\protect\JournalTitle{Annals of the University of Craiova-Mathematics and
  Computer Science Series}} \textbf{42}, 167--174 (2015).

\bibitem{gabrys2006blood}
E.~Gabry{\'s}, M.~Rybaczuk, and A.~K{\k{e}}dzia, \enquote{Blood flow simulation
  through fractal models of circulatory system,} {\protect\JournalTitle{Chaos,
  Solitons \& Fractals}} \textbf{27}, 1--7 (2006).

\bibitem{Borji2022}
A.~Borji, \enquote{{Pros and cons of GAN evaluation measures: New
  developments},} {\protect\JournalTitle{Computer Vision and Image
  Understanding}} \textbf{215}, 103329 (2022).

\bibitem{ajiboye2015evaluating}
A.~Ajiboye, R.~Abdullah-Arshah, H.~Qin, and H.~Isah-Kebbe, \enquote{{Evaluating
  the Effect of Dataset Size on Predictive Model Using},}
  {\protect\JournalTitle{Internation Journal of Software Engineering \&
  Computer Sciences}} \textbf{1}, 75--84 (2015).

\bibitem{DBLP:journals/corr/LinMBHPRDZ14}
T.~Lin, M.~Maire, S.~J. Belongie, L.~D. Bourdev, R.~B. Girshick, J.~Hays,
  P.~Perona, D.~Ramanan, P.~Doll{\'{a}}r, and C.~L. Zitnick, \enquote{Microsoft
  {COCO:} common objects in context,} {\protect\JournalTitle{CoRR}}
  \textbf{abs/1405.0312} (2014).

\end{thebibliography}

%%%%%%%%%% If preparing manually:
% \newcommand{\enquote}[1]{``#1''}

% \bibitem{Zhang:14}
% Y.~Zhang, S.~Qiao, L.~Sun, Q.~W. Shi, W.~Huang, L.~Li, and Z.~Yang,
%   \enquote{Photoinduced active terahertz metamaterials with nanostructured
%   vanadium dioxide film deposited by sol-gel method,}
%   {\protect\JournalTitle{Optics Express}} \textbf{22}, 11070--11078 (2014).

% \bibitem{Optica}
% {Optica}, \enquote{{Optica Publishing Group},}
%   \url{http://www.opg.optica.org}.

% \bibitem{FORSTER2007}
% P.~Forster, V.~Ramaswamy, P.~Artaxo, T.~Bernsten, R.~Betts, D.~Fahey,
%   J.~Haywood, J.~Lean, D.~Lowe, G.~Myhre, J.~Nganga, R.~Prinn, G.~Raga,
%   M.~Schulz, and R.~V. Dorland, \enquote{Changes in atmospheric consituents and
%   in radiative forcing,} in \enquote{Climate Change 2007: The Physical Science
%   Basis. Contribution of Working Group 1 to the Fourth assesment report of
%   Intergovernmental Panel on Climate Change,}  S.~Solomon, D.~Qin, M.~Manning,
%   Z.~Chen, M.~Marquis, K.~B. Averyt, M.~Tignor, and H.~L. Miler, eds.
%   (Cambridge University Press, 2007).

% \end{thebibliography}

\end{document}